# An Automated Tool to Detect Suicidal Susceptibility from Social Media Posts

Yasin Dus

yasindus@sentinet.org

Georgiy Nefedov

georgiy.nefedov@columbia.edu

## ABSTRACT

The World Health Organization (WHO) estimated that approximately 1.4 million individuals worldwide died by suicide in 2022. This figure indicates that one person died by suicide every 20 s during the year. Globally, suicide is the tenth-leading cause of death, while it is the second-leading cause of death among young people aged 15329 years. In 2022, it was estimated that approximately 10.5 million suicide attempts would occur. The WHO suggests that along with each completed suicide attempt, many individuals attempt suicide. Today, social media is a place in which people share their feelings. Thus, social media can help us understand the thoughts and possible actions of individuals. This study leverages this advantage and focuses on developing an automated model to use information from social media to determine whether someone is contemplating self-harm. This model is based on the Suicidal-ELECTRA model. We collected datasets of social media posts, processed them, and used them to train and fine-tune our model. Evaluation of the refined model with a testing dataset consistently yielded outstanding results. The model had an impressive accuracy rate of 93% and commendable F1 score of 0.93. Additionally, we developed an application programming interface that seamlessly integrated our tool with third-party platforms, enhancing its implementation potential to address the concern of rising suicide rates.

## Keywords

Suicide; social media; machine learning; Twitter.

## 1.INTRODUCTION

In recent years, expressing mental health concerns, such as suicidal tendencies, in social media platforms has attracted considerable attention from researchers, practitioners, and policymakers. Advancements in technology and their implications in daily life have provided individuals with novel avenues for self-expression, enabling them to share their thoughts, emotions, and experiences to digital communities [1]. The digital era has created an unprecedented opportunity to understand and address mental health issues in previously unexplored ways [2].

Mental health and suicide have been deeply linked to human existence and social structure. From ancient philosophy to modern psychiatric theories, they have been a matter of concern and are conspicuous by progress and misconceptions [3]. In contemporary society, the alarming rise in suicide rates among young individuals has signaled a need of re-evaluation of strategies aimed at prevention and support [4]. This further necessitates innovative approaches to identify those at risk and intervene in a timely manner [5]. The quest for effective prevention strategies is urgent and seeks exploration beyond conventional methodologies.

Nowadays, social media platforms have emerged as one of the primary spaces for personal expression, enabling individuals to vent their emotions and experiences to a vast audience. These digital platforms have great potential for mental health analyses, offering a unique window into the emotional states of individuals. The constantly growing volume of data shared on platforms like Twitter and Reddit presents an opportunity to understand representative patterns of mental distress, including potential signs of suicidal ideation [6, 7].

While awareness of mental health and its related challenges has improved, the ability to accurately and efficiently identify individuals at risk of suicide remains a significant research gap [8]. Current methods often rely on self-reporting or direct interaction with healthcare professionals, potentially missing those who do not actively seek help. This study aims to address this gap by leveraging the power of machine learning and natural language processing to create an automated tool that can sift through the vast digital landscape to identify those in need. The overarching goal is to develop an automated tool proficient in navigating the expansive digital landscape. Through the application of advanced computational techniques, particularly in machine learning and natural language processing, this study seeks to empower this tool to sift through the vast digital landscape with the ability to discern individuals who may require assistance or support. In essence, the aim is to create an intelligent system that can efficiently analyze digital content, identifying signals indicative of those potentially in need, thereby contributing to an automated and scalable approach for recognizing individuals requiring attention or intervention.

## 2.LITERATURE REVIEW

The present is the key to the past. Hence, it is crucial to critically examine existing research, approaches, and knowledge related to the automated detection of suicidal susceptibility from social media posts. This section aims to synthesize the current state of understanding, identify gaps, and contextualize the proposed study within the broader field of mental health analysis and technological advancements.

Analyzing suicidal behavior involves examining various psychological theories that explain the underlying factors contributing to self-destructive tendencies. The interpersonal-psychological theory of suicide proposes that an individual's capacity for suicide is influenced by the convergence of feeling left out and thinking they are a burden to others [9]. Moreover, the theory of reasoned action emphasizes how important intentions and attitudes are in predicting how we behave. This helps us understand the thoughts that influence decisions about self-harm [10].

People's actions on social media platforms can provide a lot of information about their mental health [11]. When people talk about their feelings online, they tend to use certain words and ways of expressing themselves. Researchers can investigate these online actions to find signs of mental distress that might not be understood using traditional methods [12].

Social media has become a big part of how people talk about their mental health. People use these platforms to express themselves and find support. The special aspect of social media is that it lets people from all over the world share their personal stories about mental health [13]. This is particularly helpful for noticing when someone might be struggling and determining how to help them. People have tried different ways to find posts about suicide online. Some methods use certain words, while others use algorithms to detect when someone seems really upset [14, 15]. However, these methods occasionally fail and are not very accurate. Thus, we need better ways to identify these posts quickly and correctly.

To understand and prevent suicide, many studies have been conducted on social media. In 2019, Stephen and Prabu [16], focused on how people write on Twitter to find posts that showed depression. They used sentiment analysis to identify these posts. They then used another computer program named syuzhet to determine base emotions followed by sentiment score evaluation using three different lexicons. O'dea et al. [17] employed human analysts and a computer program to investigate whether Twitter messages reflect thoughts of suicide. The study found that Twitter is indeed used by individuals to discuss suicide, and it is possible to assess the seriousness of these messages. Furthermore, researchers [18] have considered the characteristics of Twitter users who post about suicidal thoughts by analyzing their social connections rather than the content of their posts. In 2016, Kavuluru et al. [19] examined messages about suicide on Twitter. They used special computer programs to understand the meaning of these messages. These programs analyzed things like the words used, the way the messages were written, the emotions they conveyed, and the thoughts they expressed. Their goal was to differentiate between highly concerning messages, such as those indicating someone might be considering suicide, and other messages related to suicide. In 2018, researchers studied how quickly people respond to tweets related to suicide compared to those not related to suicide [20]. They discovered that responses to suicide-related tweets are notably quicker, often appearing within an hour. Studies were also conducted related to sorting tweets into two categories, one where people were talking about suicide, and another where they were not by considering the feelings expressed and using computers [21]. However, they did not proffer resolutions for instances wherein posts related to suicide could be facilitated with support or intervention to prevent self-harm.look into other ways people might talk about wanting to hurt themselves.

Birjali et al. [22] used a tool called WordNet to understand the meaning of words in tweets related to suicide. They observed numerous words specifically related to suicide, and eventually created a list of such words.

Existing research on suicide prevention often focuses on identifying thoughts of suicide or detecting posts about suicide. However, few studies specifically concentrate on identifying profiles that might be at risk of suicide.

Machine learning is a helpful tool in mental health research because it can find hidden patterns in large amounts of data. A special way of using computers to understand language, called natural language processing (NLP), has shown promising results in studying what people write to find signs of how they feel [23]. Moreover, techniques such as sentiment analysis, text classification, and clustering have been used to learn about how people express their mental health [24]. The rising implication of Natural Language Processing (NLP) has paved the way for innovative approaches [25, 26], including the utilization of LIWC [27] and other textual attributes, as well as deep learning techniques such as convolutional neural networks (CNNs), recurrent neural networks (RNNs), and bidirectional encoder representations from transformers (BERT) [28]. In the realm of suicide research, the adoption of automated Suicidality Detection plays a crucial role in suicide prevention. This method adeptly learns textual features without the need for intricate feature engineering. Some approaches incorporate these extracted features into deep neural networks (DNNs). Nobles et al. [29] applied psycholinguistic features and word occurrences in a multilayer perceptron (MLP). DNNs like convolutional neural networks (CNNs), recurrent neural networks (RNNs), and bidirectional encoder representations from transformers (BERT) are widely applied in this domain. The long short-term memory (LSTM) network, a variation of RNN, is extensively employed to encode textual sequences and subsequently classify them with fully connected layers [30]. Recent developments introduce diverse learning paradigms to collaborate with DNNs for suicide detection. Ji et al. [31] suggested model aggregation methods to update neural networks, encompassing CNNs and LSTMs, with a focus on detecting suicidal ideation in private chatting rooms. Benton et al. [32] predicted suicide attempts and mental health using neural models within a multitask learning framework that involved predicting user gender as an auxiliary task. Gaur et al. [33] enhanced text representation by incorporating external knowledge bases and a suicide-related ontology into a CNN model. Sawhney et. al. demonstrated the real-world utility of HCN, by contextualizing the impact of the conversation trees and hyperbolic learning [34]. Coppersmith et al. [35] devised a deep learning model employing GloVe for word embedding, bidirectional LSTM for sequence encoding, and a self-attention mechanism for capturing informative subsequences. Sawhney et al. [36] utilized LSTM, CNN, and RNN for Suicide Ideation Detection (SID), while Tadesse et al. [37] employed the LSTM-CNN model. Ji et al. [38] introduced an attentive relation network with LSTM and topic modeling to encode text and risk indicators. In the 2019 CLPsych Shared Task [39], various popular DNN architectures found application. Hevia et al. [40] evaluated the impact of pretraining using different models, including GRU-based RNN. Morales et al. [41] delved into several popular deep learning models, such as CNN, LSTM, and Neural Network Synthesis (NeuNetS). Matero et al. [42] presented a dual-context model utilizing hierarchically attentive RNN and BERT. This study brings together information from different areas to improve how we can find signs of suicide risk in social media posts. The upcoming sections will explain what this study wants to achieve and how it is done, analyze the collected data in detail, and discuss the findings.

## 3. OBJECTIVES OF THIS STUDY

The primary objective of this study is to develop an automated tool capable of conducting surveillance on social media platforms such as Twitter and Reddit for posts indicative of suicidal thoughts and promptly alerting platform moderators or concerned authorities for intervention and eventually, prevention. To achieve this goal, the study outlines the following specific objectives:

• Identifying suicidal posts within a large dataset acquired from Twitter and Reddit, seeking to sift through an extensive dataset of posts exhibiting linguistic patterns indicative of suicidal tendencies.

• Training and fine-tuning an existing AI model named "Suicidal-Electra" based on insights gained from analyzing the collected dataset. The existing AI model already has a proven record in identifying suicide-related posts. We anticipate that feeding more data and fine-tuning the model will not only enhance its competence but also make it more pragmatic in detecting and flagging posts warranting immediate attention.

• Developing a publicly available API such that it can be integrated with third parties easily.

In the subsequent sections, this paper demonstrates the theoretical foundation, methodological approach, data analysis, and discussion of findings. By addressing the existing gaps in the realm of mental health identification and intervention, this thesis contributes to the potential capabilities offered by technology for the prevention of suicides in the society.

## 4. MATERIALS AND METHODS

The methodology of this study constitutes data collection, data processing, model training testing, and finally, public API development and deployment.

Data acquired and utilized in this study are all data available publicly. The data procurement process is achieved by using Kaggle API. While most of the data used in this study to train, test, and validate the AI model were sourced from Twitter, Reddit-based datasets also contributed substantially. While acquiring the data was simple, processing the datasets was arduous. The entire dataset comprised social media posts from six different sources (see data sources). Although most of the dataset was already used for other research purposes, the challenge was in making the entire dataset comparable. Hence, it was crucial to clean and normalize the dataset. While cleaning the data, the first discrepancy we observed was that the labels and classes were disparate. Here, it was crucial to intervene manually as understanding the emotions from the post can be best achieved when humans read or listen to them. In the first dataset (TDAthedevastator), five data labels: "Supportive," "Ideation," "Behavior," "Attempt," and "Indicator" were used (Figure 1).

Due to its faster training and better performance on downstream tasks, we opted for the pre-trained ELECTRA [4326] model with classification head. More specifically, we fine-tune the gooohjy/suicidal-electra weights from huggingface, pre-trained on the Kaggle's "Suicide and Depression Detection" dataset.

To check which data labels were related to suicide, we manually analyzed and verified them by taking random samples of posts in different sets. Based on the samples taken, we found that the "Attempt" label clearly indicates suicide. The "Ideation" label is indicative for the state of a person when they tend toward suicidality. It is the state just before making a suicide attempt. Therefore, for this dataset, "Attempt" and "Ideation" were considered suicidal, and the rest as not suicidal.

| | User | Post | Label |
|---|---|---|---|
| 0 | user-0 | ['Its not a viable option, and youll be leavin... | Supportive |
| 1 | user-1 | ['It can be hard to appreciate the notion that... | Ideation |
| 2 | user-2 | ['Hi, so last night i was sitting on the ledge... | Behavior |
| 3 | user-3 | ['I tried to kill my self once and failed badl... | Attempt |
| 4 | user-4 | ['Hi NEM3030. What sorts of things do you enjo... | Ideation |
| ... | ... | ... | ... |
| 495 | user-495 | ['Its not the end, it just feels that way. Or ... | Supportive |
| 496 | user-496 | ['It was a skype call, but she ended it and Ve... | Indicator |
| 497 | user-497 | ['That sounds really weird.Maybe you were Dist... | Supportive |
| 498 | user-498 | ['Dont know there as dumb as it sounds I feel ... | Attempt |
| 499 | user-499 | ['>It gets better, trust me.Ive spent long ... | Behavior |

**Figure 1. Original labels assigned to posts in dataset "TDA".**

The labels in the second dataset (agAG) (Figure 2) are given both in context of suicidal or non-suicidal behavior and sentiment of the respective posts.

The suicidal labels assigned to the posts were "Suicidal" (0) and "Non-Suicidal" (1); while the assigned sentiment labels were "Negative" (0), "Positive" (1), and "Neutral" (3). Here, we explored whether the labels assigned to the posts had a significant relationship by taking different random samples and comparing them.

| | Unnamed: 0 | Post | Suicidal_label | Sentiment_label |
|---|---|---|---|---|
| 0 | 0 | Ex Wife Threatening SuicideRecently I left my ... | 0 | 0 |
| 1 | 1 | Am I weird I don t get affected by compliments... | 1 | 1 |
| 2 | 2 | Finally is almost over So I can never hear ... | 1 | 0 |
| 3 | 3 | i need helpjust help me im crying so hard | 0 | 0 |
| 4 | 4 | I m so lostHello my name is Adam and I ve b... | 0 | 0 |
| ... | ... | ... | ... | ... |
| 226948 | 227680 | I sound like a dudebro but I can t handle my f... | 0 | 0 |
| 226949 | 227681 | Fuck my sister She is such l fucking bitch and... | 1 | 0 |
| 226950 | 227682 | I ve been suicidal for years and no one knowsT... | 0 | 1 |
| 226951 | 227683 | My boyfriend is sick so I took some Polaroids ... | 1 | 0 |
| 226952 | 227684 | What would happen to my dog M What would happ... | 0 | 0 |

**Figure 2. Observed data labels in dataset "AG".**

The findings depict that based on the selected samples, the "Suicidal & Negative" labels were absolute suicidal posts. Furthermore, some "Suicidal and Positive" and "Suicidal and Neutral" labels were not completely suicidal and "sentiments" had negligible correlation with suicidality and was therefore ignored.

The data labels of the social media posts in the third (imeshsonuIMS), fourth (laxmimeritLAX), and fifth (mohanedmashalyMSH) databases are clearly expressed. While in the fifth dataset all labels are related to suicidal, in the third (imeshsonuIMS) and fourth (laxmimeritLAX) databases, only "Suicidal" and "Non-Suicidal" labels are used.

We analyzed the sixth (natalialech/ntNTLl) and final dataset by taking multiple sample sets, and our findings demonstrate five data labels (Figure 3) in this dataset, namely, "active suicidal ideation" (0), "passive suicidal ideation" (1), "sarcasm regarding suicidal ideation" (2), "suicide-related tweets" (awareness, news, chatter about suicide) (3), and other (4). After reading through multiple samples, we concluded that data labels 0 and 1 were suicidal, whereas labels 2 and 4 were non-suicidal. However, keeping in mind the objectives of this study, we discarded the news data (Label 3), as this data may create conflict in training in the identification of suicidality. A detailed outlook of the number of registers for each label in each dataset is presented in Figure 4.

In processing the datasets, making the data comparable is necessary before the data is used to train and fine-tune the Suicidal-Electra model. Therefore, we merged all data to a single dataset having two classes: non-suicidal (0) and suicidal (1). Subsequently, it is important to clean, make corrections, and normalize all textual contents in the data. Text normalization is an essential pre-processing step that improves the quality of the text and makes it suitable for machines to process. When we normalize text, we try to make it similar to a usual form to help the computer handle less diverse information, which makes it work better. We simplified the textual contents and made it machine-ready for the model by executing a recursive summarizer.

**Figure 3. Demonstration of presence of six different labels in social media posts in dataset "NTL".**

Recursive summarizer summarizes textual content of 300 or more words while ensuring that the meaning remains unchanged. Posts with less than 300 words are restricted from recursive summarizer applications. The minimum and maximum length of the derived summary was set between 50 and 120 words for this study. Any potential bias that could arise due to large textual content was removed after the dataset was merged and before the recursive summarizer was applied. We analyzed the distribution of the cleaned data and found that 95% of the textual contents in the data fall within a count of 430 words (Figure 5). For this study, textual contents with lengths greater than 430 words are considered outliers and removed from the dataset.

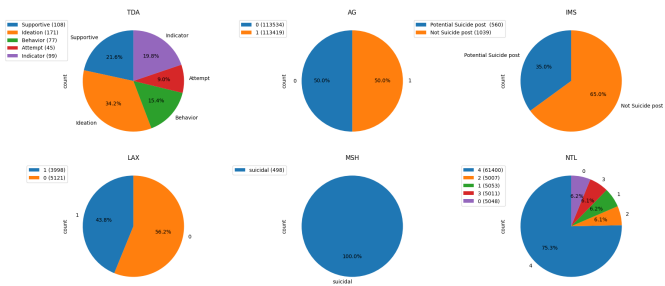

**Figure 4. Comprehensive breakdown of register counts for each label in all datasets.**

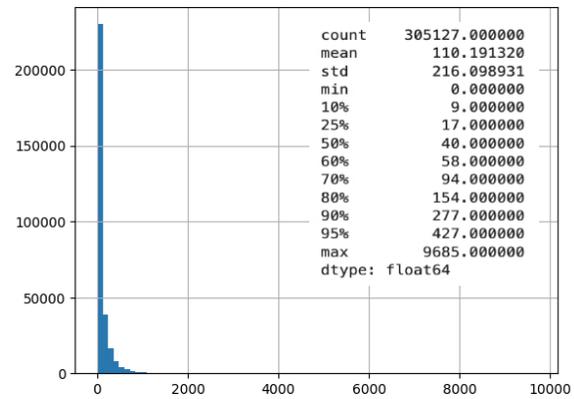

**Figure 5. Distribution of posts in accordance with word counts.**

As per the methods implemented to prepare the data for "Suicidal-Electra" model, we implemented a number of steps to prepare the dataset. These steps include removing accented characters from text, extra whitespaces, URLs, symbols, digits, special characters, fix spellings, and word lengthening (characters are wrongly repeated). In the next steps, we tokenized the text and omitted stop words. In the subsequent phase of pre-processing, we performed techniques such as stemming and lemmatization to simplify words. A sample of the cleaned, summarized, and normalized dataset is given in Figure 6.

**Figure 6. Sample data after application of pre-processing and text normalization.**

Stemming involves simplifying words by getting rid of extra parts and turning them into their root form. The goal is to group similar words, even if the root form is not a word in the dictionary. For example, words such as "connection," "connected," and "connecting" all become the common word "connect." Lemmatization, in contrast, changes words to their main forms, handling different forms correctly and ensuring they are proper words in the language. It is a more advanced process compared to stemming, which deals with individual words without considering the bigger picture. In lemmatization, the root word is called a "lemma." This word serves as the standard dictionary or reference form for a group of words. On pre-processing the datasets, suicidal and non-suicidal labels and classes were assigned to the processed data. The distribution shows that the datasets consist of more non-suicidal posts in comparison to suicidal posts (Figure 7).

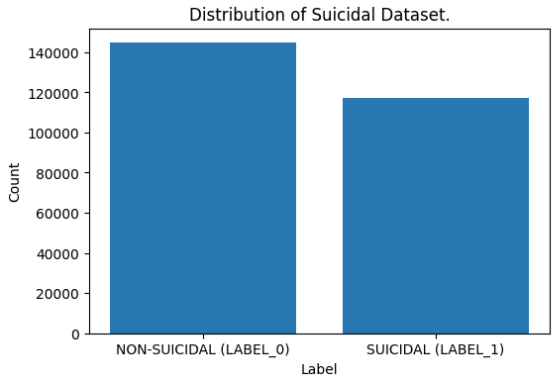

**Figure 7. Distribution of suicidal and non-suicidal labeled textual classes**

We further checked the most frequent words (Figure 8) manually. This attempt was to ensure that no unrelated words are present for predication that can change the accuracy of the model.

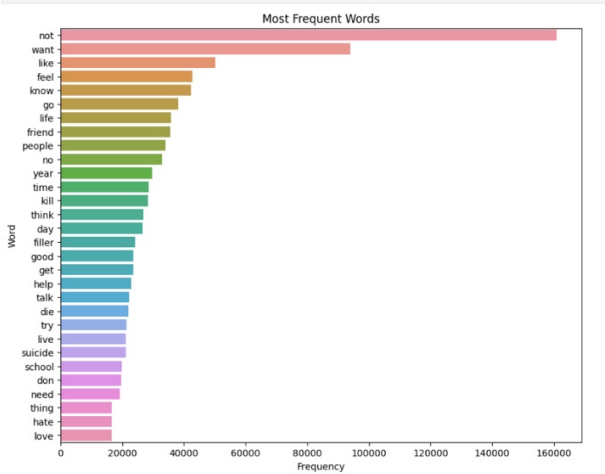

**Figure 8. Occurrence of the topmost frequent words in the dataset.**

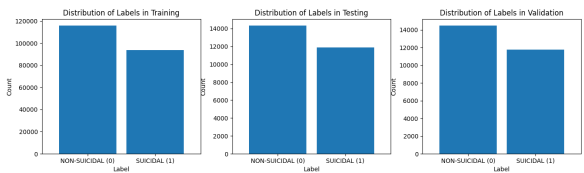

**Figure 9. Analysis of dataset distribution, detailing the counts for non-suicidal and suicidal classes among training, testing, and validation sets.**

In this study, we utilized an already existing text classification model for suicide prediction, "Suicidal-Electra." Suicidal ELECTRA, is fine-tuned based on the latest ELECTRA transformer model, known for achieving state-of-the-art performance in classifying suicidal text. Although we didn't directly compare it with other baseline models in this study, providing more context is important. During the model building and selection process of the, Suicidal ELECTRA [44] based on which the fine-tuned model of this study is derived, various models evaluated, including Logistic Regression (Logit), Convolutional Neural Network (CNN), Long Short-term Memory (LSTM), Bidirectional Encoder Representations from Transformers (BERT), and Efficiently Learning on Encoder that Classifies Token Replacements Accurately (ELECTRA). These models were trained on the training dataset, fine-tuned on the validation dataset, and tested on the test dataset. Evaluation metrics such as Accuracy, Precision, Recall, and F1 score were employed to assess their performance on the test dataset. Given the specific focus on the F1 score for our use case, which aims to predict suicide or non-suicide, it a more balanced perspective of model performance as compared to recall. The key results of the models executed are given in Table 1. The comparison revealed that transformer models, especially ELECTRA, outperformed other models across all metrics. ELECTRA achieved the highest F1 score, accuracy, and recall, while BERT exhibited the highest precision. Despite the comparable performance of these two models, ELECTRA demonstrated superior speed and effectiveness, addressing the limitations associated with BERT's use of MLM. Consequently, ELECTRA was chosen for detecting suicidality in social media posts and was utilized in this study.

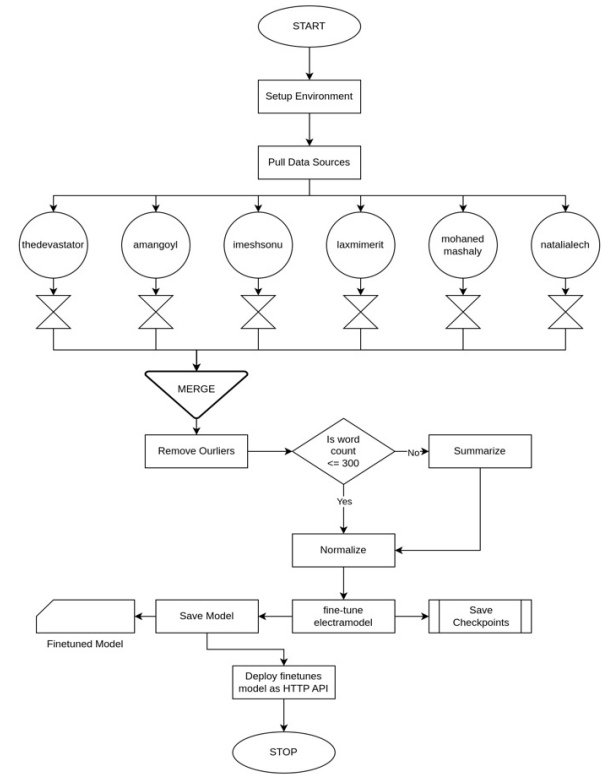

**Figure 8. Schematic representation of the processes and steps implemented in this study.**

Suicidal ElectraIt is engineered to determine suicidal or non-suicidal tendencies from a given sequence of words. At its core, "Suicidal-Electra" rests upon a foundation of deep learning and NLP. Its training encompasses a vast corpus of annotated text samples of both suicidal and non-suicidal expressions. By processing and learning from these examples, the model acquires a solid understanding of linguistic patterns associated with suicidal thoughts. This training equips the model with a deep understanding of language patterns, metaphors, and expressions often associated with feelings of despair. What sets this model apart is its exceptional ability to comprehend the subtle distinctions of human language and context. Architecturally, this model is rooted in transformer-based neural networks with competence in apprehending contextual relationships within sequences of data. Within this architecture, self-attention mechanisms facilitate the model's ability to weigh the significance of different words based on their contextual relevance. ELECTRA introduced a novel pre-training task known as Replaced Token Detection (RTD) to address the limitations of the MLM technique employed by BERT, which involves corrupting input with masked tokens. Unlike BERT, which learns to predict only a small subset of tokens (the masked tokens) rather than predicting every input token, RTD ensures a more comprehensive learning process for each sentence. Table 2 depicts the architecture on which this model is based. Overall, "Suicidal-Electra" is an example of how AI can understand language. Its transformer structure figures out the meaning of words in a sentence, effectively identifying signs that someone might be thinking about suicide. This model shows how powerful machine learning can be in dealing with important issues in society. The detailed architecture, working algorithm, its training dataset etc. are thoroughly discussed in Suicidal Text Detection Report, 2022 [25]. Fine-tuning "Suicidal-Electra" on more specific datasets can harness its potential to substantially impact online mental health support. Fine-tuning the model with more datasets will customize it, enabling it to recognize subtle signs unique to certain communities, age groups, or cultural contexts. The model could thus become an invaluable asset for varied users.

| Best Model | Accuracy | Recall | Precision | F1 Score |
|---|---|---|---|---|
| Logit* | 0.9111 | 0.8870 | 0.8832 | 0.8851 |
| CNN* | 0.9285 | 0.9013 | 0.9125 | 0.9069 |
| LSTM* | 0.9260 | 0.8649 | 0.9386 | 0.9003 |
| BERT# | 0.9757 | 0.9669 | 0.9701 | 0.9685 |
| ELECTRA# | 0.9792 | 0.9788 | 0.9677 | 0.9732 |

**Table 1: Comparison of Electra with other existing models.**

This study aimed to fine-tune "Suicidal-Electra" with the expectation of enhancing the model's ability to accurately detect suicidal tendencies; further, we intended to contribute to a safer and more supportive online environment. This model can save lives by identifying distress signals that could otherwise be disregarded. In short, the advanced language comprehension and adaptability of "Suicidal-Electra" make it the best option to detect suicidal tendencies from social media posts.

| Parameter | Value |
|---|---|
| **Attention Dropout Probability** The dropout probability for attention probabilities. | 0.1 |
| **Embedding Size** The size of the embeddings for each token in the input sequence. | 768 |
| **Hidden Activation** The activation function used in the hidden layers. | gelu (Gaussian Error Linear Unit) |
| **Hidden Dropout Probability** The dropout probability for hidden layers. | 0.1 |
| **Hidden Size** The size of the hidden layers. | 768 |
| **Intermediate Size** The size of the intermediate (feed-forward) layer in the transformer. | 3072 |
| **Max Position Embeddings** The maximum sequence length the model can handle. | 512 |
| **Number of Attention Heads** The number of attention heads in the multi-head attention models. In this case, there are 12 attention heads. | 12 |
| **Number of Hidden Layers** The number of hidden layers in the transformer. | 12 |
| **Summary Activation** The activation function used in the summary layer. | gelu |
| **Summary Last Dropout** The dropout probability for the summary layer. | 0.1 |
| **Vocabulary Size** The size of the vocabulary for input tokens. | 30522 |

**Table 2: Insights of Architecture of Suicidal Electra model.**

In the ensuing pre-processing steps, we trained "Suicidal-Electra." We followed the training guidelines of the original training data for the said model. In this course, the entire data is distributed for training, validation, and testing in 80%, 10%, and 10% respectively. (Figure 9). The breakdown of the dataset distribution including counts for non-suicidal and suicidal classes in the training are 116147, 93,707; in testing are 14,355, 11,877, and in validation sets are 14,491, 11,741 respectively. Upon fine-tuning the model, we saved the model, which finally deployed as HTTP API. A visual representation of the entire process from data acquisition to API deployment is given in Figure 10.

In operational terms, the developed API deploys NLP, intensively probing the text to detect subtle indicators of suicidal thoughts. The seamless integration of this tool with online platforms can facilitate prompt and effective interventions that could assist individuals in need. Rapid interventions can make a substantive difference and offer a lifeline to individuals who could be struggling silently. This API could equip social media platforms with the capacity to extend a helping hand and create a safer space in which individuals feel heard, understood, and supported.

Furthermore, incorporating this tool into social media platforms would denote a proactive step toward building caring communities. Such an amalgamation can represent a platform's commitment to its users' welfare. Users engaging with such a social media platform can trust that it transcends surface-level interactions, watches out for them, and is prepared to help if distress signals are detected.

Overall, the deployment of this API can foster safe and compassionate online environments. Social media platforms can evolve into havens of support by recognizing signs of suicidality using advanced NLP. Such online forums can demonstrate a profound understanding of the real-world challenges confronting individuals.

## 5. RESULTS

In our diligent process of evaluation of the Suicidal-Electra model's performance on the classification task, we employed a comprehensive analysis using the confusion matrix (Figure 11). This matrix provides a detailed breakdown of the model's predictive accuracy, including True Positives (TP: correctly identified suicidal posts), False Positives (FP: non-suicidal posts mistakenly classified as suicidal), True Negatives (TN: correctly identified non-suicidal posts), and False Negatives (FN: suicidal posts mistakenly classified as non-suicidal). The resulting values of 13643, 841, 707, and 11011 for TP, FP, TN, and FN respectively, showcase the model's ability to make a fine distinction in the challenging task of detecting suicidal susceptibility in social media posts.

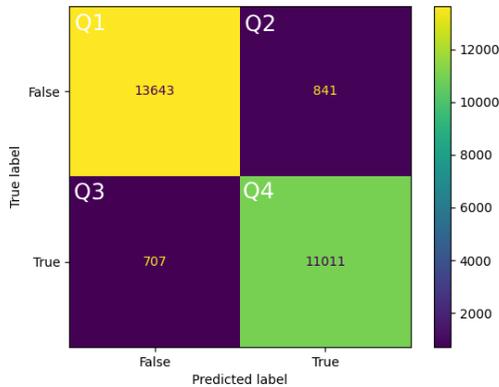

**Figure 11. A detailed analysis of the confusion matrix depicting in-depth divisions of the models' predictive accuracy.**

The accuracy rate of 93% underscores the model's overall effectiveness (Table 3), indicating the proportion of correct predictions among the total predictions made. Furthermore, the F1 score of 0.93 is a significant metric, especially for a task as sensitive as identifying suicidal content. This score represents a harmonic mean of precision and recall, highlighting the model's capability to strike a balance between minimizing false positives and false negatives.

| RESULTS | |
|---|---|
| Accuracy | 0.939432 |
| Recall | 0,937164 |
| Precision | 0,928222 |
| F1 | 0,932672 |

**Table 3. Metrics of fine-tuned model for evaluation.**

## 6. DISCUSSION

Results underscore the effectiveness of our model in accurately detecting and classifying suicidal susceptibility in social media posts. The high F1 score indicates a strong balance between precision and recall, which is commendable for such a sensitive task. Upon evaluating the fine-tuned model with the testing dataset, we observed consistently excellent outcomes. The model demonstrates a remarkable ability to identify suicidal posts with a high degree of confidence. In a significant proportion of tests, the model's confidence level stands at approximately 0.99 (Figure 12), further substantiating its robust performance.

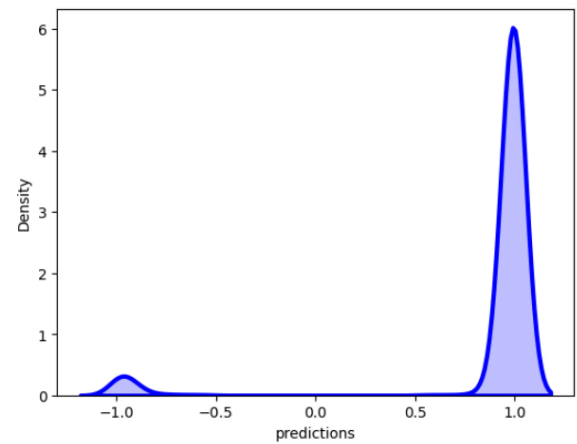

**Figure 12. Depiction of the accuracy and the model's ability to identify suicidal posts.**

We conducted a comprehensive comparison between our model's predictions and the actual outcomes, particularly in terms of the success and failure rates. The efficiency of our model is striking, with a minimal 6.45% of predictions turning out to be incorrect. This showcases the model's ability to make accurate determinations regarding the presence of suicidal content in social media posts. Figure 13 presents a visual representation of the prediction distribution. This graphical depiction provides insights into the distribution of successful and unsuccessful predictions made by our model.

**Figure 13. Density plot of successful and unsuccessful predictions of the fine-tuned model. In this plot, +1 and −1 indicates successful and failed predictions respectively.**

The overall results affirm the efficacy of our automated tool in identifying and detecting suicidal susceptibility in social media posts. The high accuracy, confident predictions, and efficient performance underscore the potential of the model in making a meaningful contribution to the field of mental health support through digital platforms.

As we move forward, we intend to further refine and enhance the model's capabilities, striving for even greater accuracy and robustness in our automated detection tool. Your valuable insights and feedback would be instrumental in shaping the future direction of our research.

The current study focuses on establishing this tool as a primary detector for suicide-related posts to enable faster intervention. We acknowledge the potential introduction of biases during model training with the utilized datasets, given the dataset's size used for fine-tuning is not particularly extensive. For future technical enhancements, we aim to improve data quality by employing semi-supervised learning techniques for pseudo labelling. Using a small set of human-annotated data to train deep neural networks in a supervised manner, we intend to combine human-labelled and pseudo-labelled data to enhance the training of our models for suicidal prediction. This strategy aims to improve the overall robustness of the model by ensuring higher data quality. While pseudo-labelled data may be slightly less accurate than human-annotated data, its significant advantage lies in reducing the manual hours required for labelling. Additionally, in this study, we utilized the Suicidal ELECTRA model on our dataset, as computational constraints prevented the integration of the ELECTRA large model. Larger transformer models, pre-trained with more data, layers, and parameters, have demonstrated superior results on benchmark datasets. However, it's important to note that the utilization of larger transformer models for improved performance comes with the trade-off of longer training times and the potential risk of overfitting on the training dataset.

## 7. CONCLUSIONS

In conclusion, our study has focused on developing an automated tool to identify potential signs of suicidal thoughts in social media posts from social media platforms like Twitter and Reddit. We successfully categorized the entire dataset into either suicidal or non-suicidal classes. Training and fine-tuning the "Suicidal-Electra" model, has significantly enhanced our ability to understand and pinpoint posts with potential suicidal content. Notably, the model's performance has been assessed and verified to be effective. The obtained results have been promising, demonstrating an impressive accuracy of 93%. This achievement underscores the potential of AI-driven analysis in detecting suicidal susceptibility from social media expressions. Currently, this model is fine-tuned.

Moving forward, our future endeavors aim at refining the accuracy of suicidal post detection. This involves deeper investigation, evaluating the varying degrees of suicidality within posts, as well as identifying profiles at risk with greater precision. Social media platforms, such as Facebook and Instagram, will be a key aspect of our forthcoming research. This extension will enable us to comprehensively address the broader landscape of social media content and contribute to a comprehensive understanding of mental health issues in the digital realm.

44, 2 (June 2020), 933–955. https://doi.org/10.25300/MISQ/2020/14110

15. Jorge Lopez-Castroman, Bilel Moulahi, Jérôme Azé, Sandra Bringay, Julie Deninotti, Sebastien Guillaume, and Enrique Baca-Garcia. 2020. Mining social networks to improve suicide prevention: A scoping review. Journal of Neuroscience Research, 98, 4 (April 2020), 616–625. https://doi.org/10.1002/jnr.24404

16. Jini J. Stephen and Prabu P. 2019. Detecting the magnitude of depression in Twitter users using sentiment analysis. International Journal of Electrical and Computer Engineering, 9, 4 (August 2019), 3247. https://doi.org/10.11591/ijece.v9i4.pp3247-3255

17. Bridianne O'Dea, Stephen Wan, Philip J. Batterham, Alison L. Calear, Cecile Paris, and Helen Christensen. 2015. Detecting suicidality on Twitter. Internet Interventions 2, 2 (May 2015), 183–188. https://doi.org/10.1016/j.invent.2015.03.005

18. Gualtiero B. Colombo, Pete Burnap, Andrei Hodorog, and Jonathan Scourfield. 2016. Analysing the connectivity and communication of suicidal users on twitter. Computer Communications 73, (January 2016), 291–300. https://doi.org/10.1016/j.comcom.2015.07.018

19. Ramakanth Kavuluru, María Ramos-Morales, Tara Holaday, Amanda G. Williams, Laura Haye, and Julie Cerel. 2016. Classification of helpful comments on online suicide watch forums. In Proceedings of the 7th ACM International Conference on Bioinformatics, Computational Biology, and Health Informatics (BCB '16), Association for Computing Machinery, New York, NY, USA, 32–40. https://doi.org/10.1145/2975167.2975170

20. Bridianne O'Dea, Melinda R. Achilles, Mark E. Larsen, Philip J. Batterham, Alison L. Calear, and Helen Christensen. 2018. The rate of reply and nature of responses to suicide-related posts on Twitter. Internet Interventions 13, (September 2018), 105–107. https://doi.org/10.1016/j.invent.2018.07.004

21. Scott R Braithwaite, Christophe Giraud-Carrier, Josh West, Michael D Barnes, and Carl Lee Hanson. 2016. Validating machine learning algorithms for Twitter data against established measures of suicidality. JMIR Mental Health 3, 2 (May 2016), e21. https://doi.org/10.2196/mental.4822

22. Marouane Birjali, Abderrahim Beni-Hssane, and Mohammed Erritali. 2017. Machine learning and semantic sentiment analysis based algorithms for suicide sentiment prediction in social networks. Procedia Computer Science 113, (January 2017), 65–72. https://doi.org/10.1016/j.procs.2017.08.290

23. Anupama B. S, Rakshith D. B, Rahul Kumar M, and Navaneeth M. 2020. Real time Twitter sentiment analysis using natural language processing. International Journal of Engineering Research & Technology 9, 7 (July 2020). https://doi.org/10.17577/IJERTV9IS070406

24. Ferdaous Benrouba and Rachid Boudour. 2023. Emotional sentiment analysis of social media content for mental health safety. Social Network Analysis & Mining 13, 1 (January 2023), 17. https://doi.org/10.1007/s13278-022-01000-9

25. Aeron S. S. Win, Goh J. Yi, Lim Z. Hui, Lin Xiao, and Quek Y. Zhen. 2022. Suicidal text detection in social media posts. Dissertation. BT4222 Mining Web Data for Business Insights. Retrieved from https://github.com/gohjiayi/suicidal-text-detection/blob/main/docs/Suicidal-Text-Detection_Report.pdf

26. Kevin Clark, Minh-Thang Luong, Quoc V. Lee, Christopher D. Manning. ELECTRA: Pre-training Text Encoders as Discriminators Rather Than Generators
## 9. DATA SOURCES

https://www.kaggle.com/datasets/thedevastator/c-ssrs-labeled-suicidality-in-500-anonymized-red

https://www.kaggle.com/datasets/amangoyl/reddit-dataset-for-multi-task-nlp

https://www.kaggle.com/datasets/imeshsonu/suicideal-phrases

https://raw.githubusercontent.com/laxmimerit/twitter-suicidal-intention-dataset/master/twitter-suicidal_data.csv

https://www.kaggle.com/datasets/mohanedmashaly/suicide-notes

https://www.kaggle.com/datasets/natalialech/suicidal-ideation-on-twitter